\documentclass[prd,preprint,showpacs,preprintnumbers,amsmath,amssymb,nofootinbib]{revtex4-1}

\usepackage{graphicx}
\usepackage{dcolumn}
\usepackage{bm}
\usepackage{amsfonts}
\usepackage{CJK}

\def\fm {\mathop{\hbox{fm}}}
\def\MeV {\mathop{\hbox{MeV}}}

\def\Re {\mathop{\hbox{Re}}}

\def\Tr {\mathop{\hbox{Tr}}}
\def\DU  {\mathop{{\cal D}\hbox{U}}}
\def\Dpsi {\mathop{{\cal D}\bar{\psi}{\cal D}\psi}}
\def\dd  {\mbox{d}}
\newcommand\detn[1]{\mbox{det}_{#1}}

\newcommand\bdetn[1]{\overline{\mbox{det}}_{#1}}

\newcommand{\beq}{\begin{equation}}
\newcommand{\eeq}{\end{equation}}
\newcommand{\beqa}{\begin{eqnarray}}
\newcommand{\eeqa}{\end{eqnarray}}

\newcommand\comment[1]{}

\begin{document}

\preprint{UK/10-04}

\begin{CJK*}{UTF8}{} 
\title{Finite density phase transition of QCD with $N_f=4$ and $N_f=2$ using canonical
ensemble method}
\CJKfamily{gbsn}
\author{Anyi Li (李安意)}
\email{anyili@phy.duke.edu}
 \affiliation{Department of Physics and Astronomy,
University of Kentucky, Lexington, Kentucky 40506, USA \\
Department of Physics, Duke University, Durham, North Carolina 27708, USA}
\author{Andrei Alexandru}
 \affiliation{Physics Department, The George Washington University, Washington DC 20052, USA}

\author{Keh-Fei Liu (刘克非)} 
 \affiliation{Department of Physics and Astronomy,
University of Kentucky, Lexington, Kentucky 40506, USA}
\CJKfamily{gbsn}
\author{Xiangfei Meng (孟祥飞)}

\affiliation{Department of
Physics, Nankai University, Tianjin 300071, China\\
National Supercomputing Center, Tianjin 300457, China}
\collaboration{$\chi$QCD Collaboration}


\begin{abstract}
In a progress toward searching for the QCD critical
point, we study the finite density phase transition of $N_f = 4$ and 2
lattice QCD at finite temperature with the canonical ensemble approach.
We develop a winding number expansion method to accurately project out the
particle number from the fermion determinant which greatly extends the
applicable range of baryon number sectors to make the study feasible.
Our lattice simulation was carried out with
the clover fermions and improved gauge action. For a given temperature,
we calculate the baryon chemical potential from the canonical approach to look for the
mixed phase as a signal for the first order phase transition. In the case
of $N_f=4$, we observe an ``S-shape'' structure in the chemical potential-density
plane due to the surface tension of the mixed phase in a finite volume which is
a signal for the first order phase transition. We use the Maxwell construction to
determine the phase boundaries for three temperatures below $T_c$. The intersecting point of the two extrapolated boundaries turns out to be at the expected first order transition point at $T_c$ with $\mu = 0$. This serves as a check for our method of identifying the critical point.
We also studied the $N_f =2$ case, but do not see a signal of the mixed phase for temperature as low as 0.83 $T_c$.
\end{abstract}


\maketitle
\end{CJK*}

\section{Introduction}

\begin{figure}[h!]
  \center
  \includegraphics[scale=0.30]{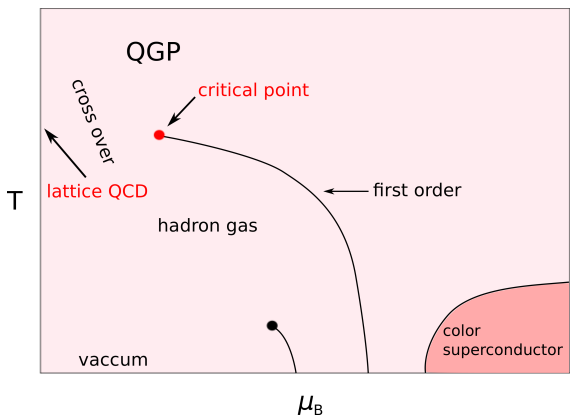}\\
  \caption[A conjectured QCD phase diagram]{Conjectured QCD phase diagram}\label{fig:qcd-phase-diagram}
\end{figure}

Quantum Chromodynamics (QCD) is a fundamental theory which
describes the strong interaction of quarks and gluons, from which the nucleons are made of. The
knowledge of QCD at finite temperature and finite baryon chemical potential is essential to
the understanding of a variety of phenomena. Exploring the QCD phase diagram, including identification of 
different phases and
the determination of the phase transition line is currently one of the intensely studied 
topics~\cite{Karsch:2003jg, Muroya:2003qs}.
Guided by phenomenology and experiments, many candidate phase diagrams have been proposed, one of such conjectured phase diagrams~\cite{Rajagopal:2000wf} is sketched in  
Fig.~\ref{fig:qcd-phase-diagram}. In certain limits, in
particular for large temperatures $T$ or large baryon-chemical potential $\mu_B$, thermodynamics is dominated by 
short-distance QCD dynamics and the theory can be studied analytically. However, the most interesting phenomena lie 
in regions where non-perturbative features of the
theory dominate. The only known systematic approach is the first-principle calculation via lattice QCD  using
Monte Carlo simulations.

The thermodynamics of strongly interacting matter has been studied extensively in lattice calculations at vanishing baryon
chemical potential. Current
lattice calculations strongly suggest that the transition from the low temperature hadronic phase to
the high temperature phase is a continuous but rapid crossover happening in a narrow temperature interval around $T_c
\sim 170 \MeV$~\cite{Aoki:2006we,Cheng:2006qk,Detar:2007as,Karsch:2007dt,Karsch:2008fe,Aoki:2006br,Aoki:2009sc}. On
the other part of the phase diagram~---~large baryon chemical potential but very low temperature, a number of different model
approaches~\cite{Alford:1998mk,Brown:1990ev,Asakawa:1989bq,Barducci:1989wi,Barducci:1989eu,Barducci:1993bh,Berges:1998rc,Halasz:1998qr}
suggest that the transition in this region is strongly first order, although this argument is less robust.
This first order phase transition is expected to become less pronounced as we lower the chemical potential and it
should terminate at a second order phase transition point -- the critical point.

The search for the QCD critical point has attracted considerable theoretical and experimental
attention recently. The possible existence of this point was
introduced sometime ago~\cite{Asakawa:1989bq,Barducci:1989wi}. It is apparent that the location of the
critical point is a key to understanding QCD phase diagram. For experimental search of the critical point, it has been
proposed to use heavy ion collisions at RHIC~\cite{Stephanov:1998dy,Stephanov:1999zu}. The appearance of this point is closely
related to hadronic fluctuations which may be examined by an
event-by-event analysis of experimental results. The upcoming RHIC lower energy scan and FAIR (GSI) will focus on
the region $T > 100 \MeV$ and $\mu_B \sim 600 \MeV$ where the critical point is predicted to exist in theoretical models.
However, in order to extract unambiguous signals for the QCD critical
point, quantitative calculations from first-principle lattice QCD are indispensable.

Remarkable progress in lattice simulations at zero baryon chemical potential has been made
in recent years; however, simulations at non-zero chemical
potential are difficult due to the complex nature of the fermionic determinant at non-zero chemical potential. The
phase fluctuations produce the notorious ``sign problem''.
The majority of current simulations are focusing on the small chemical potential region $\mu_q/T \ll 1$ where the ``sign problem''
appears to be controllable. Most of them are based on the grand canonical ensemble ($T$, $\mu_B$ as parameters).
Since the existence and location of
the critical point is still unknown, we need an algorithm which can be applied beyond the small
chemical potential region. This is one of the motivations for our studies via the canonical ensemble
approach~\cite{Engels:1999tz, Liu:2000dj, Liu:2002qr,Azcoiti:2004ri,deForcrand:2006ec}.

We have proposed an algorithm based on the canonical partition function to alleviate the overlap and
fluctuation problems~\cite{Liu:2003wy,Alexandru:2005ix}.
The method we use is computationally expensive since
every update involves the evaluation of the fermionic
determinant; however, finite baryon density simulation based on this method has been shown to be
feasible at temperature above $\sim 80$\% $T_c$ where the sign problem is under control~\cite{Alexandru:2005ix,Alexandru:2007bb}. In this approach, we measure the baryon chemical potential 
to detect the phase transition.
With the aid of the winding number expansion technique~\cite{Meng:2008hj,Danzer:2009sr,Gattringer:2009wi,Danzer:2008xs},
a program was outlined to
scan the QCD phase diagram in an effort to look for the critical point~\cite{Li:2006qa,Li:2007bj}.

In the present work, we shall consider the cases with four- and two-flavors of degenerate
quarks. It is numerically easier to carry out HMC simulations with even number of flavors. Furthermore, 
the four-flavor case is known to have a first order phase transition at finite temperature with $\mu=0$ 
and the phase transition line is extended to the finite $\mu$. We shall use the Maxwell construction to 
determine the phase boundaries of the mixed phase in the canonical ensemble approach at several temperatures 
lower than $T_c$ and extrapolate the
boundaries to see if they meet at $T_c$ and $\mu=0$. This provides us a desirable
testing ground for our method before we launch a more numerically intensive simulation 
to search for the critical point in the $N_f = 3$ case. Besides, we can apply the method to
the two-flavor case which could be similar to the three-flavor case and, in this case, will
likely have a critical point which we can attempt to locate. 

In this paper, we present results based on simulations on $6^3\times 4$ lattices with clover fermion action for  
two and four flavors QCD with quark masses which correspond to pion masses at $\sim 700 - 800$ MeV.
We plot the chemical potential as a function of baryon density. In finite volume, due to the non-zero contribution 
from the surface tension, the first order phase transition will be signaled by an ``S-shape'' structure in this plot. 
The phase boundaries of the coexistent phase can be determined by ``Maxwell construction''. 
We clearly observe an S-shape structure in the simulation of the $N_f=4$ case, indicating a first order phase 
transition~\cite{deForcrand:2006ec}. In our present study, we do not see such a structure in the $N_f = 2$ case down to 0.83 $T_c$.

The paper is organized as follows: canonical partition function method
is outlined in \mbox{Sec. II}, we also present winding 
number expansion methods (WNEM) and simulation parameters in
this section. In Sec. III, we  carry out the measurement of the baryon
chemical potential as an attempt to determine the phase boundaries in order to scan the phase diagrams with $N_f=4$ and $N_f=2$. 
We then conclude our study in Sec IV and 
give an outline for the future simulation
of $N_f=3$ case.

\section{Algorithm}

\subsection{Canonical partition function}

The canonical partition function in lattice QCD can be derived from the fugacity expansion of the grand canonical partition function,
\beq
Z(V,T,\mu) = \sum_{k} Z_C(V, T, k) e^{\mu k/T}, \label{eq:fugacity}
\eeq
where $k$ is the net number of quarks (number of quarks minus the number of anti-quarks) and $Z_C$ is the canonical 
partition function. We note here that
on a finite lattice, the maximum net number of quarks is limited by the Pauli exclusion principle. Using the fugacity expansion, 
it can be shown that the canonical partition function can be written as a Fourier transform of the grand canonical partition function,
\beq   \label{eq:Z_C}
Z_C(V, T, k) =
\frac{1}{2\pi} \int_0^{2\pi} \mbox{d}\phi \,e^{-i k \phi} Z(V, T,\mu)|_{\mu=i\phi T},
\eeq
after introducing an imaginary chemical potential $\mu = i \phi T$.

For our study, we will focus on Iwasaki improved gauge action and clover fermion action from the following papers~\cite{Iwasaki:1996ya,AliKhan:2000iz,AliKhan:2001ek}. They are defined as
\beqa
Z(\beta, \kappa , \mu) &=& \int
\DU(\det M(U))^{N_f} e^{-S_g(U)},
\\
S_g(U) &=& -\beta \left\{ \sum_{x,\, \mu > \nu}
      c_{0} W_{\mu \nu}^{1 \times 1}(x)(U)
    + \sum_{x,\, \mu, \nu} c_{1} W_{\mu \nu}^{1 \times 2}(x)(U) \right\},
\\
  M_{x,y}(U) &=& \delta_{x,y}
  - \delta_{x,y} c_{sw} \kappa \sum_{\mu > \nu} \sigma_{\mu \nu} F_{\mu \nu}
    -\kappa \sum_{i} \left[(1-\gamma_i) U_i(x)~\delta_{x+\hat{i},y}
    +(1+\gamma_i){U_i}^\dagger(x-\hat{i})~\delta_{x-\hat{i},y}\right]
     \nonumber \\
  && -\kappa \left[e^{-\mu} (1-\gamma_4) U_4(x)~\delta_{x+\hat{4},y}
    +e^{\mu} (1+\gamma_4) {U_4}^\dagger(x-\hat{4})~\delta_{x-\hat{4},y}
    \right],
\eeqa
where $W_{\mu \nu}^{1 \times 1}(x)$ and $W_{\mu \nu}^{1 \times 2}(x)$ are $1 \times 1$ and $1 \times 2$ Wilson loops, 
$F_{\mu \nu}=(f_{\mu \nu} - f_{\mu
\nu}^\dagger)/(8i)$, where $f_{\mu \nu}$ is the standard clover-shaped combination of gauge links, $\beta=6/g^2$, $c_1=-0.331$, 
$c_0=1-8c_1$. We adopt
nonperturbative $O(a)$ improved $c_{sw}$\cite{Aoki:2005et}, $c_{sw} = 1 + 0.113 (\frac{6}{\beta}) + 0.0158 (\frac{6}{\beta})^2 + 0.0088
(\frac{6}{\beta})^3$ for $N_f=2$ and $c_{sw}= 1 + 0.113 (\frac{6}{\beta}) + 0.0209 (\frac{6}{\beta})^2 + 0.0047 (\frac{6}{\beta})^3$ 
for $N_f = 4$ case.
$\mu \equiv \mu_q a$ is the quark chemical potential which is introduced at every temporal link. 
One can perform a change of 
variables~\cite{Engels:1999tz},
\beqa
\psi(\vec{x},x_4)&\rightarrow& \psi'(\vec{x},x_4)=
e^{-\mu x_4}\psi(\vec{x},x_4) \nonumber \\
\bar{\psi}(\vec{x},x_4)&\rightarrow& \bar{\psi}'(\vec{x},x_4)=
e^{\mu x_4}\bar{\psi}(\vec{x},x_4)
\eeqa
to absorb $\mu$ in the fermionic field. 
This leaves the chemical potential to reside only on the the time-direction links  on the last time slice.  
Now the part of the fermion matrix $M(x,y,\mu/T)$ which involves the chemical potential becomes
\begin{equation}  \label{eq:mu/T}
M(x,y,\mu)= -\kappa \left[e^{-\mu N_{t}} (1-\gamma_4) U_4(\vec{x},N_t)~\delta_{x_4,N_t}\delta_{y_4,1}
    +e^{\mu N_{t}} (1+\gamma_4) {U_4}^\dagger(\vec{x},N_t)~\delta_{x_4,1}\delta_{y_4,N_t} \right]~\delta_{\vec{x},\vec{y}}.
\end{equation}
In terms of the imaginary chemical potential, it corresponds to a $U(1)$ phase attached to the
last time slice
\beq
(U_\phi)_\nu(x)\equiv\left\{
\begin{array}{ll}
U_\nu(x) e^{-i\phi} & x_4=N_t ,\, \nu=4 \\
U_\nu(x) & \mbox{otherwise},
\end{array}
\right.
\eeq
where $\mu N_t $ in Eq.~(\ref{eq:mu/T}) is replaced with $i\phi$.

After integrating out the fermionic part in Eq.~(\ref{eq:Z_C}), we get an expression
\beq
Z_C(V, T, k) = \int \DU e^{-S_g(U)} \detn{k} \label{eq:canonical}
M^{N_f}(U),
\eeq
where
\beq
\detn{k} M^{N_f}(U) \equiv \frac{1}{2\pi}\int_0^{2\pi} \dd\phi\,e^{-i k \phi} \det M(m,
\phi;U)^{N_f} , \label{eq:detk}
\eeq is the  projected determinant with the fixed net quark number $k$. 

We shall summarize some of the properties of the canonical ensemble:

\begin{itemize}
\item

    Since the fermion Dirac matrix with the $U(1)$ phase on one time slice still satisfies 
$\gamma_5$ hermiticity, 
\beq
 \gamma_5 M(U,\phi)\gamma_5 = M^{\dagger}(U,\phi), 
\eeq
the determinant $\det M(U,\phi) = \det M(U,\phi)^*$ is real. 

\item

     From the charge conjugation property of the Wilson-clover fermion matrix
\beq
M(U,\phi) = C^{-1} M(U^c, -\phi)^{T}C,
\eeq
where $U^c = U^*$ is the link variable under charge conjugation and $C$ is the gamma matrix
which has the transformation property $C^{-1} \gamma_{\mu} C = - \gamma_{\mu}^{T}$
and the fact that ${U}$ and ${U^c}$ configurations has equal weight in the path integral, one 
can show that
\beq  \label{eq:ZCeven}
Z_C(V,T,k) = Z_C(V,T, -k) .
\eeq
In other words, the canonical partition function for $k$ quarks is equal to that of
$k$ anti-quarks. Since the canonical partition function is even in $k$, 
Eq.~(\ref{eq:canonical}) becomes 
\beq  \label{eq:det_cos}
Z_C(V, T, k) = \int \DU e^{-S_g(U)} {\rm Re\,}\detn{k} M^{N_f}(U),
\eeq 
 and $Z_C(V,T,k)$ is real, but not necessarily positive. Due to the Fourier transform, 
there can be a sign problem at large quark number and low temperature.
Eq.~(\ref{eq:ZCeven}) can be similarly derived from considering time reversal transformation.
Alternatively, one can take the charge-$\gamma_5$ hermiticity transform (CH) property of
the fermion matrix
\beq
M(U,\phi) = C^{-1}\gamma_5 M^*(U^c, -\phi)\gamma_5C,
\eeq
and the fact that the gauge action in invariant under charge conjugation, i.e. $S_g(U) = S_g(U^c)$ and 
that ${U}$ and ${U^c}$ configurations has equal weight in the path integral to show that
the partition function $Z_C(V,T,k)$ is real. 
\comment{
Thus, one comes to the conclusion that
the projected determinant involves $\cos(k\phi)$ as in Eq.~(\ref{eq:det_cos}) and that
Eq.~(\ref{eq:ZCeven}) holds.
}

\item

    Fermion operators will involve projections in different quark sectors. For example,
a quark bilinear $\overline{\psi}\Gamma\psi\equiv \sum_{f=1}^{N_{f}}\overline{\psi}_{f}\Gamma\psi_{f}$ has the following expression for its
expectation value in the canonical ensemble with $k$ quarks~\cite{Alexandru:2005ix}
\beqa
\left< \bar{\psi}\Gamma\psi\right>_{k} &=& \frac{1}{Z_C(k)}\frac{1}{2\pi}
\int_{0}^{2\pi} {\rm d}\phi e^{-ik\phi} \int \DU e^{-S_g(U)} \nonumber \\
&\times&\int \Dpsi e^{-S_f(U_{\phi},\bar{\psi},\psi)}
\bar{\psi}\Gamma\psi  \\
&=& \left\langle\sum_{k'} \frac{\detn{k'}M^{N_f}}{\detn{k}M^{N_f}}
(-N_f\mbox{Tr}_{k-k'}\Gamma M^{-1})\right\rangle, \nonumber
\eeqa
where we have considered the Fourier transform from Eq.~(\ref{eq:detk}) and we define
\beq   \label{eq:detprojdisc}
\mbox{Tr}_k \Gamma M^{-1} \equiv \frac{1}{2\pi}\int_{0}^{2\pi}{\rm d}\phi e^{-ik\phi}
\Tr \Gamma M(U_{\phi})^{-1}.
\eeq
The summation above runs over all integer values of $k$ but the Fourier coefficients are non-zero
only for $-6 V_{s}\leq k \leq6 V_{s}$, with $V_{s}$ the spatial volume in
lattice units. As it is clear from the convolution in quark
number, sectors with different $k$ will contribute to fermionic observables.

\item

The quark number can be evaluated with the conserved vector charge,
\beqa
J_4 = -\kappa \sum_{f=1}^{N_{f}}\sum_{\vec{x}} [\bar{\psi}_{f}(x+\hat{t})(1+\gamma_4)U_4^\dagger(x)\psi(x)_{f} \nonumber \\
-\bar{\psi}_{f}(x)(1-\gamma_4)U_4(x)\psi_{f}(x+\hat{t})].
\eeqa
\beq
\langle J_4\rangle_k = \frac{1}{Z_C(k)}\int \DU e^{-S_g(U)} 
\frac{1}{2\pi}\int_0^{2\pi} \dd\phi\,e^{-i k \phi} \det M(U,\phi)^{N_f} (-N_f) \Tr[O],
\eeq
where $O = -\kappa [(1+\gamma_4)U_4^\dagger M^{-1}(U,\phi) - (1-\gamma_4)U_4 M^{-1}(U,\phi)] $. 
From $\gamma_5$ hermiticity, one can show that
\beq
\Tr[O] = -\Tr[O^*],
\eeq
which means that it is imaginary. On the other hand, from CH transformation, we
find that $\langle J_4\rangle_k$ is real. Therefore,
\beq   \label{eq:J4}
\langle J_4\rangle_k = \frac{1}{Z_C(k)}\int \DU e^{-S_g(U)} 
\frac{1}{2\pi}\int_0^{2\pi} \dd\phi\,(-i)\sin(k \phi) \det M(U,\phi)^{N_f} (-N_f) \Tr[O],
\eeq
which is odd in $k$ as it should for a charge-odd operator. 

\comment{
It has been shown in an earlier test with discrete Fourier transform, 
that the quark number is indeed $k$ from the canonical ensemble for $k$ with explicit
calculation in Eq.~(\ref{eq:J4}). However, }
Using the equation above, it is straightforward to show that
\beqa
\langle n\rangle_k &=& \langle J_4\rangle_k \nonumber \\
 &=&\frac{1}{Z_C(k)}\int \DU e^{-S_g(U)} 
\frac{1}{2\pi}\int_0^{2\pi} \dd\phi\,e^{-i k \phi} \frac{d}{d\phi}(\det M(U,\phi)^{N_f}) =k
\eeqa

\comment{
This is in contrast to the grand canonical ensemble with isospin chemical potential where
$\langle n\rangle = 0$.
}

\item 
   For quark numbers which are multiple of 3, the canonical partition function is invariant
under the transformation where all the temporal links in a time-slice are multiplied 
with a $Z_{3}$ element. 
In this simulation, we shall consider the triality zero sector with integral baryon numbers 
$k = 3 n_{B}$ by implementing a $Z_3$ hopping in the Hybrid Monte Carlo simulation~\cite{Kratochvila:2003,Alexandru:2005ix} 
to avoid being frozen in one $Z_3$ sector. 

\end{itemize}

To simulate Eq.~(\ref{eq:canonical}) dynamically, we can rewrite canonical partition function as
\beqa
  Z_C(V, T, k) &= &\int \DU e^{-S_g(U)} \detn{k} M^{N_f}(U)
  \nonumber \\
  &=&\int \DU e^{-S_g(U)} \mbox{det}M^{N_f}(U)W(U)\alpha(U),
\eeqa
where
\beq
W(U) = \frac{\mathop{|\Re}\detn{k}M^{N_f}(U)|}{\mbox{det}M^{N_f}(U)}
\eeq
and
\beq
\alpha(U) = \frac{\detn{k}M^{N_f}(U)}{|\mathop{\Re}\detn{k}M^{N_f}(U)|}\label{eq:sign_def}
\eeq
Our strategy to generate an ensemble is to employ Metropolis accept/reject method based on 
weight $W(U)$ and fold the phase factor 
$\alpha(U)$ into the
measurements. In short, during the simulation, the candidate configuration is ``proposed'' by the standard Hybrid Monte Carol
algorithm\cite{Gottlieb:1987mq,Duane:1987de} and then an accept/reject step is used to correct the probability. 

We note that the 
accept/reject step is
designed to be based on the determinant
ratio which has been shown to alleviate the fluctuation problem~\cite{Joo:2001bz,Liu:2002qr,Liu:2003wy} 
and enhance the acceptance rate.

\subsection{Winding number expansion}

The majority of the time in the simulation is 
spent on the accept/reject step, specifically on computing the determinant of the fermion matrix. On the $6^3\times4$ lattice,
the matrix has $10368 \times 10368$ entries. Although the matrix is very sparse, 
exact determinant calculation is very demanding even on 
this small lattice. An alternative is to use a noisy
estimator~\cite{Alexandru:2007bb,Joo:2001bz}. In this study we use an exact evaluation of 
the determinant.
One technical challenge has to do with the Fourier transform; our original approach was to use an approximation where 
the continuous integration is replaced with a 
discrete sum, i.e.
\begin{equation}
\detn{k} M^{N_f}(U) \approx \frac{1}{N} \sum_{j=0}^{N-1} e^{-i k
\phi_j} \det M(U_{\phi_j})^{N_f},~~~~~\phi_j=\frac{2\pi
j}{N}.\label{eq:fourier}
\end{equation}

\begin{figure}[h]
\centering
\includegraphics[scale=0.55]{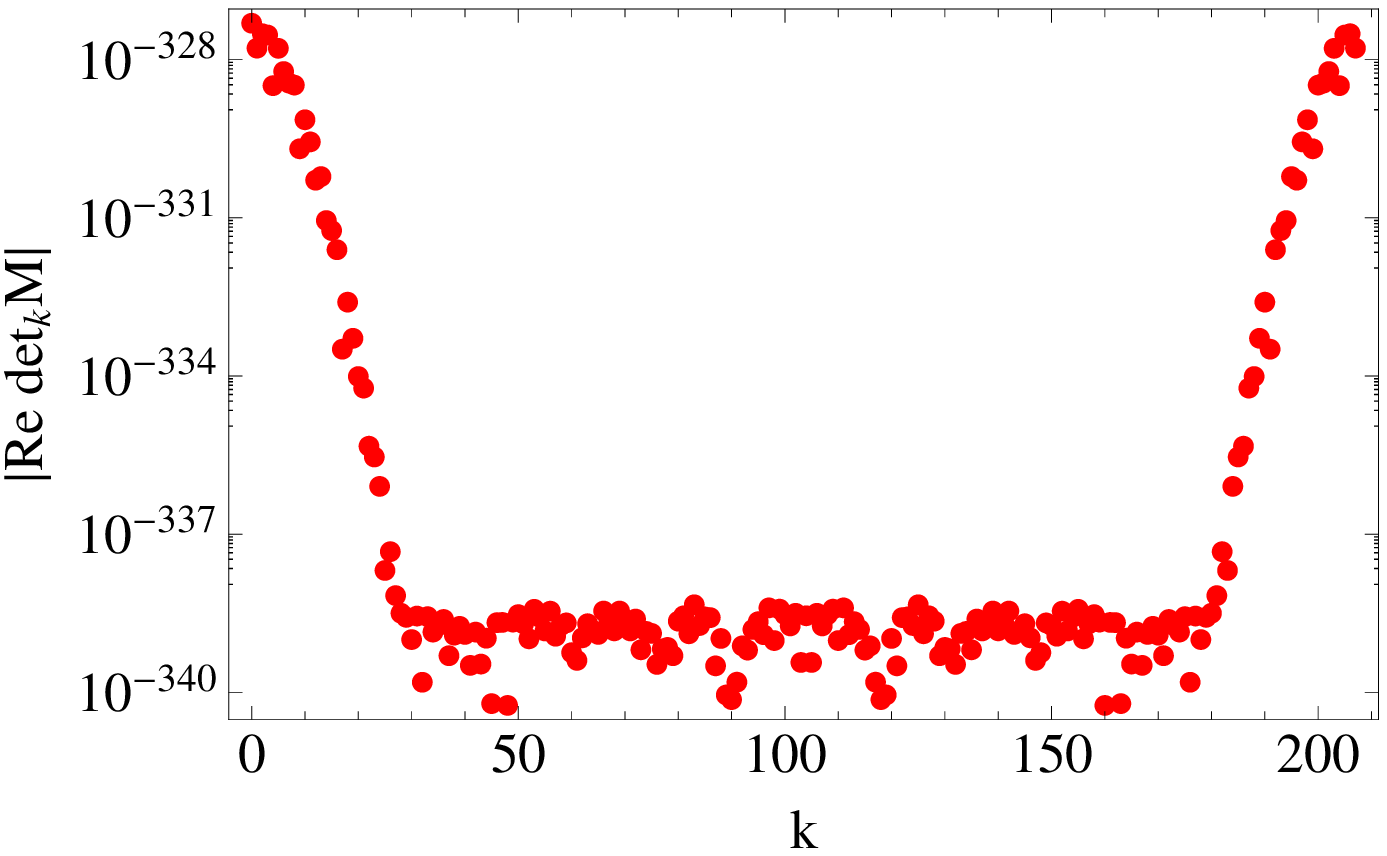}
\includegraphics[scale=0.55]{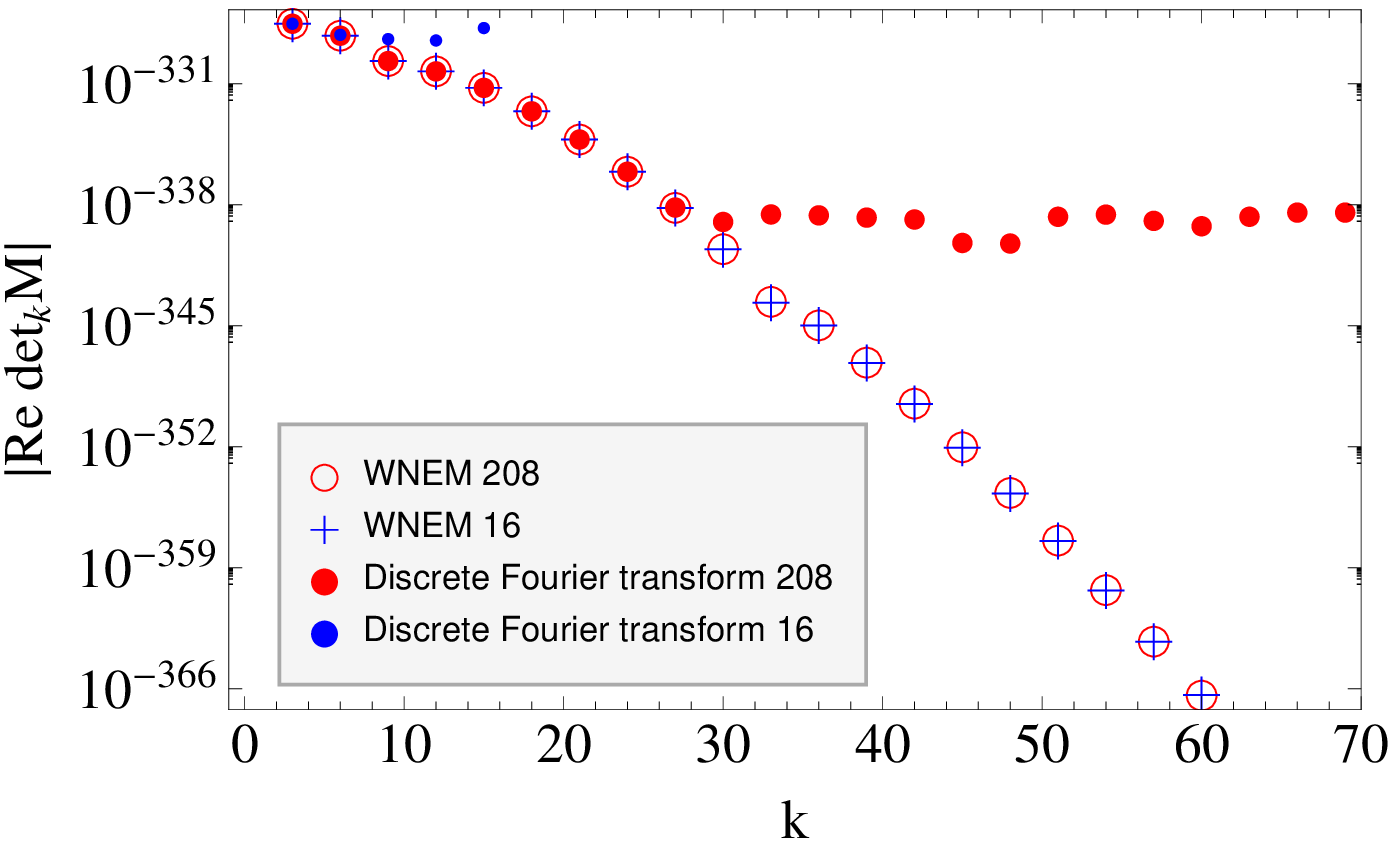}
\caption{Numerical instability of discrete Fourier transform
with 208 points (Left). Comparison from winding number expansion method and discrete Fourier transform with $N=16$ 
evaluations (Right). 
\label{fig:discrete}}
\end{figure}

It was shown that the errors introduced by this approximation are small~\cite{Li:2007bj} for small quark numbers. 
However, there are two problems
with this approach: (i) the computation time increases linearly with the net quark number,
(ii) although evaluating the determinant $N$ times should theoretically allow us to compute projection number as large as
$k=N/2$, numerically we found that for large quark numbers, the Fourier components become too
small to be evaluated with enough precision, even
using double precision floating point numbers. It has been shown~\cite{Meng:2008hj} that the results of the projected determinant
for $k$ larger than $20$ would differ significantly for different choice of $N$, which signals a numerical instability.
To see this problem, we use $N=208$
to evaluate the fermion determinant and calculate Fourier projection using discrete Fourier
transform. One would expect  $|\mathop{\Re}\detn{k}M(U)|$ to decrease exponentially since it is proportional to the free energy.
We see that this is indeed
the case for discrete Fourier transform at small quark number, as shown in
Fig.~\ref{fig:discrete} (left panel). However, as the quark number gets close to 30, the
projected determinant calculated using the discrete Fourier transform
flattens out when it reaches the double precision limit of $10^{-15}$.
This is the onset of numerical instability.

This happens because the Fourier coefficient decreases very fast as the quark number
increases and it quickly becomes comparable to machine precision.
The accumulation of round-off errors makes it impossible to evaluate the projected determinant.
This numerical challenge led us to develop a new method which
should be free of the above numerical problem for quark numbers relevant to the phase transition
region in this study~\cite{Meng:2008hj}.
The idea of our new method is to first consider the Fourier transform of $\log \det M(U,\phi)$ instead of
the direct Fourier transform of $\det M(U,\phi)$. Using an
approximation based on the first few components of $\log \det M(U,\phi)$, we can then analytically
compute the projected determinant. The efficacy of the
method can be traced to the fact that the Fourier components of $\log \det M(U,\phi)$ are the number of terms in the 
expansion which characterizes the number of quark loops wrapping around the time boundary. They are exponentially smaller 
with increasing winding numbers. This is why we can
approximate the exponent of the determinant very accurately with a few terms which, in turn, allows us to evaluate 
the Fourier components of the determinant precisely.

To see how this works, we look at the hopping expansion of $\log\det M(U,\phi)$. We start by
writing the determinant in terms of the trace log of the quark matrix
\beq
\label{eq:trlog}
\det M(U,\phi)=\exp(\log\det M(U,\phi)) = \exp(\Tr\log M(U,\phi)).
\eeq

It is well known that $\Tr\log M$ corresponds to a sum of quark loops. We classify these loops in terms of the number of times they wrap around the lattice in the temporal direction. Since the determinant is real, the loop
expansion is 
\beqa
\label{eq:quarkloop}
 &\Tr \log M(U,\phi)&=\sum_{loops}L(U,\phi)\\{\nonumber}
 &&= A_{0}(U)+
 [\sum_{n}e^{in\phi}W_{n}(U)+e^{-in\phi}W_{n}^{\dag}(U)],
\eeqa
where $n$ is the net winding number of the quark
loops wrapping around the time direction and $W_{n}$ is
the weight associated with all these loops with winding number $n$. $A_0(U)$ is the contribution with zero winding number. 

\begin{figure}[hbt!]
\centering
\includegraphics[scale=0.45]{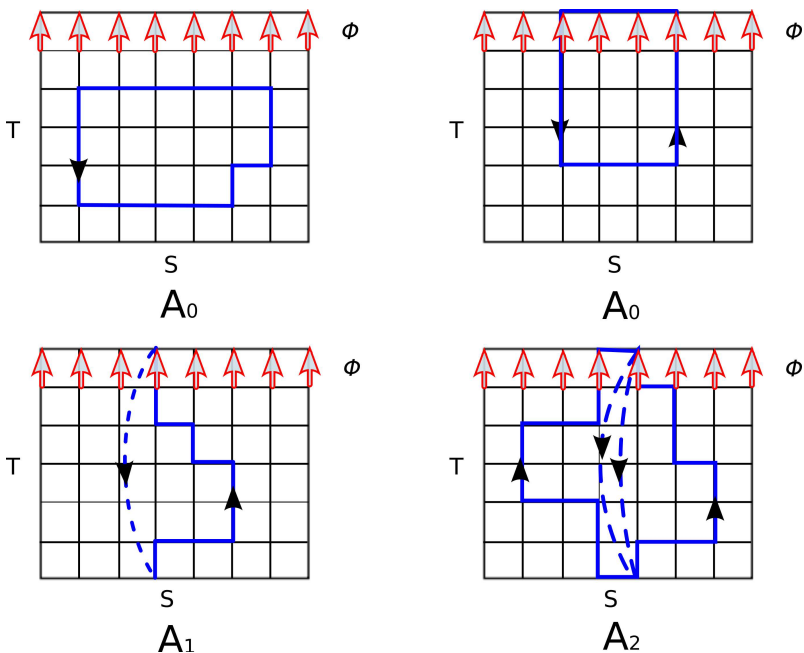}
\caption{Demonstration of first few winding loops in the winding number expansion. 
The arrows at the top of each plot show where the U(1) phases reside. 
The top two plots show quark loops which do not 
pick up the $U(1)$ phase, so they only contribute to the $\phi$ 
independent term $A_0$. 
The lower left and lower right plots show quark loops wrapping around the lattice
once and twice in the time direction which will contribute to $A_1$ and $A_2$ respectively.}
\label{fig:winding_loop}
\end{figure}

Eq.~(\ref{eq:quarkloop}) can also be re-written as
\beqa
\label{eq:wne expression}
 &\Tr \log M(U,\phi)&= A_{0}(U)+
 [\sum_{n}e^{in\phi}W_{n}(U)+e^{-in\phi}W_{n}^{\dag}(U)]\\{\nonumber}
 &&=A_{0}(U)+\sum_{n} A_{n}
 \cos(n\phi+\delta_{n}),
\eeqa
where $A_n \equiv 2|W_n|$ and $\delta_n \equiv \arg (W_n)$ are independent of $\phi$. The pictorial contribution of the first few $A_n(U)$ are demonstrated in Fig.~\ref{fig:winding_loop}.
Using Eq.~(\ref{eq:trlog}) and Eq.~(\ref{eq:wne expression}) we get
\beq
\label{eq:logdet wne} \det
M(U,\phi)=e^{A_{0}+A_{1}\cos(\phi+\delta_{1})+A_{2}\cos(2\phi+\delta_{2})+.....}.
\eeq

The Fourier transform of the first order in the expansion can now be computed analytically.
\beq
\int_{0}^{2\pi}\frac{d\phi}{2\pi}e^{-ik\phi}e^{A_0+A_{1}\cos(\phi+\delta_{1})}=e^{A_0+ik\delta_{1}}{
I}_k(A_1),
\eeq where ${I}_k$ is the Bessel function of the first kind with rank $k$.

For higher orders in the winding number expansion, we compute the
Fourier transform based on the truncated Taylor expansion
\begin{eqnarray}
&&\int_{0}^{2\pi}\frac{d\phi}{2\pi}e^{-ik\phi}e^{A_0+A_{1}\cos(\phi+\delta_{1})}
e^{\sum_{k=2}^{\infty}A_{k}\cos(k\phi+\delta_{k})}
{\nonumber}\\
&=&\int_{0}^{2\pi}\frac{d\phi}{2\pi}e^{-ik\phi}e^{A_0+A_{1}\cos(\phi+\delta_{1})}
\prod_{n=2}^{\infty} \sum_{m_n=0}^{\infty}\frac{A_n^{m_n}}{m_n!} \cos(n\phi+\delta_n)^{m_n} \nonumber\\
&=&c_{00}{\mathop{\hbox{I}}}{_{k}}(A_{1})+c_{+01}{\mathop{\hbox{I}}}{_{k+1}}(A_{1})+c_{-01}{\mathop{\hbox{I}}}{_{k-1}}(A_{1})+c_{+02}{\mathop{\hbox{I}}}{_
{k+2}}(A_{1})+...
\label{eq:final wne}
\end{eqnarray}
The projected determinant is written in terms of the linear combination of Bessel functions,
the coefficients $c$ can be easily computed
analytically. Using Eq.~\eqref{eq:final wne} and the recursion relation for the Bessel function,
${\mathop{\hbox{I}}}{_{k-1}}(A)=\frac{2k}{A}{\mathop{\hbox{I}}}{_{k}}(A)+{\mathop{\hbox{I}}}{_{k+1}}(A)$,
the winding number expansion method (WNEM) can be extended to higher orders.

For the purpose of generating configurations we use an approximation where we
keep only 6 winding loops in the WNEM expansion for $\Tr \log M$. To compute the projected determinant
we use the above Taylor expansion where we keep the first 10 orders in Taylor expansion for the second and third term in WNEM
and only the first order for the higher winding number terms. Referring to Eq.~\eqref{eq:final wne} this means
that we keep only terms with $n\leq 6$, $m_{2,3}\leq 10$ and $m_{4,5,6}\leq 1$. 
We tuned these parameters by comparing the 
results of this approximation with a high order approximation (which we regard as exact to machine
precision) on a set of gauge configurations that were generated in a previous study. In the right panel 
of Fig.~\ref{fig:discrete} we pick a particular configuration and compare the results of this approximation, 
labeled as WNEM 16, with the high order approximation, labeled WNEM 208. It can be seen that
in the range of quark 
number of interest ($k\leq40$) the approximation is very accurate. This approximation only
requires 16 evaluations of the determinant for each accept/reject step; we will denote the value of the
projected determinant computed this way with $\bdetn{k}M$.

This approximation is successful because the higher order terms contribute little to the
value of the projected determinant. It is possible that the approximation is less accurate on configurations 
generated using a different set of parametes than the ones we used to tune the approximation. To
address this concern, we correct the possible errors when
we compute the observables. All we have to do is to use a slightly different reweighting factor; we
replace the factor in Eq.~\eqref{eq:sign_def} with
\beq
\alpha(U) = \frac{\detn{k}M(U)}{|\mathop{\Re}\bdetn{k}M(U)|}.
\label{eq:sign_rew}
\eeq
This removes any bias the approximation might introduce. 
The only possible issue is to make sure that the reweighting phase doesn't introduce
too much noise in the observable. This can be gauged by determining how much the average ratio 
$|\Re \detn{k}M|/|\Re \bdetn{k}M|$ differs from 1. In our simulations we find that this ratio always stays close
to one.

To compute the projected determinant $\detn{k} M$ to machine precision 
we use a high order WNEM approximation and numerical integration of the Fourier transform. We note
that the numerical integration requires using precision higher than machine precision; we use GMP 
library and perform the Fourier transform
using 512-bit precision. The high order WNEM approximation involves computing the
determinant for many phases (64-128) for each configuration in our ensembles. Ideally, we would like to 
use this method to generate configurations but this would cost 4-8 times more computational resources.  
On the other hand, to correct the approximation bias we only have to evaluate this on the saved configurations 
which represent a small fraction of the configurations proposed during the dynamical evolution.

Before we conclude this section, we would like to compare the merits of WNEM with those of discrete Fourier transform. We compute the values of the projected determinant using 
the discrete Fourier transform with $N=16$ in Eq.~\eqref{eq:fourier}.
In Fig.~\ref{fig:discrete} (right panel), these results are compared with those from a WNEM approximation
using the same number of determinant evaluations.
We see that the discrete Fourier transform is only valid up to $k=6$.
This is to be compared to WNEM (16) which takes
the same computational time and yet does not suffer from this
problem and, as expected, the evaluated projected determinant continues to decrease as the quark number is increased.
It is this projected determinant evaluated from WNEM
that allows us to scan a wide range of densities in the QCD phase diagram.

\subsection{Simulation parameters}

A major part of the time in these simulations
is spent on computing the fermion matrix determinant. We
compute the determinant using a numerical package, SuperLU~\cite{lidemmel03}, which speeds up 
sparse matrix determinant calculation by a factor of 10 when compared to the usual dense matrix 
routine from LAPACK. In this study, we work on $6^3 \times 4$ lattices with quark masses which correspond to
pion masses at $700-800$ MeV. Since the theoretical complexity of this algorithm is between $O(n^2)$ and $O(n^3)$,
the exact determinant calculation will be very time consuming for larger matrices. Even with the SuperLU package, the high computational cost prevents us from investigating bigger lattices at the present stage. 

The Fourier transform involves the evaluation of the fermion matrix determinant $N$ times.
Thus the computation cost increase linearly with $N$. With the aid the WNEM,
$N=16$ is found to provide precise enough results for current settings. For this study we used a machine which has $16$ cores per node. We optimize our
algorithm to distribute the $16$ determinant calculations with one determinant calculation per core,
so that the calculation of each determinant is done in a parallel fashion. 
Comparing to multi-core parallel computation of 16 determinants sequentially, 
we save almost $40\%$ of the computer time.

We fixed the scale by using $r_0$~\cite{Sommer:1993ce} and the pion mass is
determined using dynamically generated ensembles at zero
temperature on $12^3\times 24$ lattices for different $\beta$.
To locate the pseudo critical temperature $T_c$ at zero chemical potential, we varied
$\beta$ to look for the peak of the Polyakov loop susceptibility.
We fixed $\kappa$ at an intermediate quark mass value ($\kappa=0.1371$) in order to avoid the unphysical
parity-broken phase, the``Aoki phase''~\cite{Aoki:1983qi,Aoki:1986xr,Aoki:1989rw,Aoki:1990ap,Aoki:1992nb,Aoki:1996af}.
Since our volume in lattice units is small and quark mass is heavy,
we present our results in terms of the temperature ratio $T/T_c$ rather than converting
them to physical units.
The temperature $T/T_c$ ratio is determined by
measuring the lattice spacing $a(\beta,\kappa)$ at zero-temperature as
\beq
\frac{T}{T_c}(\beta,\kappa)=\frac{N_t\times a(\beta_c,\kappa)}{N_t\times a(\beta,\kappa)}
\eeq

\begin{table}[!ht]
  \centering
  \caption{Simulation parameters for $N_f=2$}\label{tb:param_f2}
  \begin{tabular}{|c|c|c|c|c|}
  \hline
  $\beta$ & $a(\fm)$ & $m_\pi(\MeV)$ & $V^{-1}(\fm^{-3})$ & $T/T_c$\\
  \hline
  1.75 & 0.321(2) & 710(10) & 0.139(6) & 0.83(1) \\
  1.77 & 0.306(1) & 744(8) & 0.162(4) &  0.87(2)\\
  1.79 & 0.291(2) & 744(11) & 0.188(5) & 0.92(1)\\
  1.81 & 0.270(3) & 719(14) & 0.235(19) & 1\\
  \hline
\end{tabular}
\end{table}

\begin{table}[!ht]
  \centering
  \caption{Simulation parameters for $N_f=4$}\label{tb:param_f4}
  \begin{tabular}{|c|c|c|c|c|}
  \hline
  $\beta$ & $a(\fm)$ & $m_\pi(\MeV)$ & $V^{-1}(\fm^{-3})$ & $T/T_c$\\
  \hline
  1.56 & 0.351(2) & 831(10) & 0.107(7) & 0.89(1) \\
  1.58 & 0.340(5) & 828(10) & 0.118(5) & 0.91(2)\\
  1.60 & 0.328(2) & 834(8) & 0.132(12) & 0.95(2)\\
  1.62 & 0.312(4) & 848(6) & 0.152(14) & 1\\
  \hline
\end{tabular}
\end{table}

We list the simulation parameters ($\beta$, lattice spacing $a$, $m_{\pi}$, inverse spatial volume,
and $T/T_c$) in Table~\ref{tb:param_f2} for $N_f =2$ and Table~\ref{tb:param_f4} for
$N_f = 4$.

From these tables, we note that the pion mass varies very
little with $\beta$, consequently the quark mass is roughly the
same in all runs. We also note that the
quark mass is quite heavy, around the strange quark mass.
We chose simulation temperatures below the
pseudo-critical temperature because the phase transition at finite chemical potential is believed to
appear below the transition point at zero temperature.

To determine the location of the phase transition at non-zero baryon density, we pick
three temperatures and scan the baryon density while monitoring the chemical potential.
The ``S-shape'' wiggle in the chemical potential plot signals the presence of the phase transition.
We vary the density by changing the net quark number and scan the density range between
4 and 20 times the nuclear matter density.

For the HMC proposal step, we adopted the $\phi$ algorithm~\cite{Gottlieb:1987mq} made exact
by an accept/reject step at the end of each
trajectory\cite{Duane:1987de}. For the updating process, we set the length of the trajectories
to 0.5 with $\delta\tau=0.01$. The HMC acceptance rate was very
close to 1 since the step length is
small. We adjust the number of HMC trajectories between two consecutive finite density accept/reject
steps so that the acceptance rate at different $k$
stays in the range of 15\% to 30\%.

\section{Results}

\subsection{Sign problem}
\begin{figure}[hbtp!]
\centering
      \includegraphics[scale=0.55]{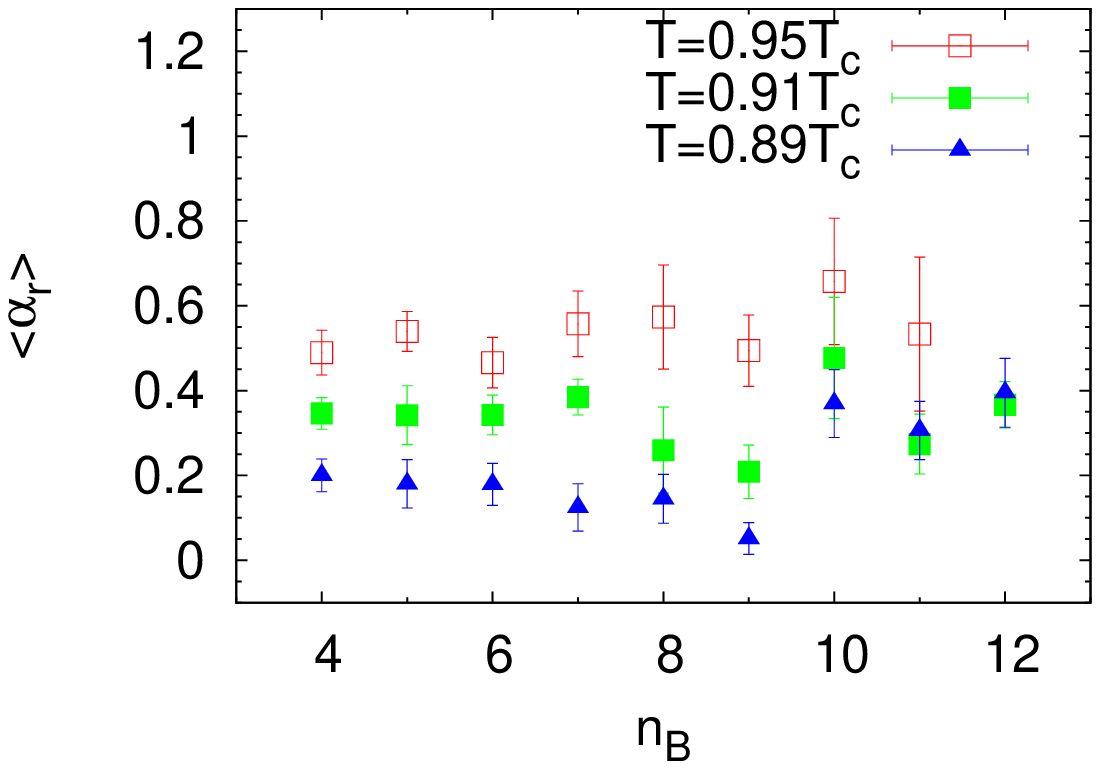}
      \includegraphics[scale=0.55]{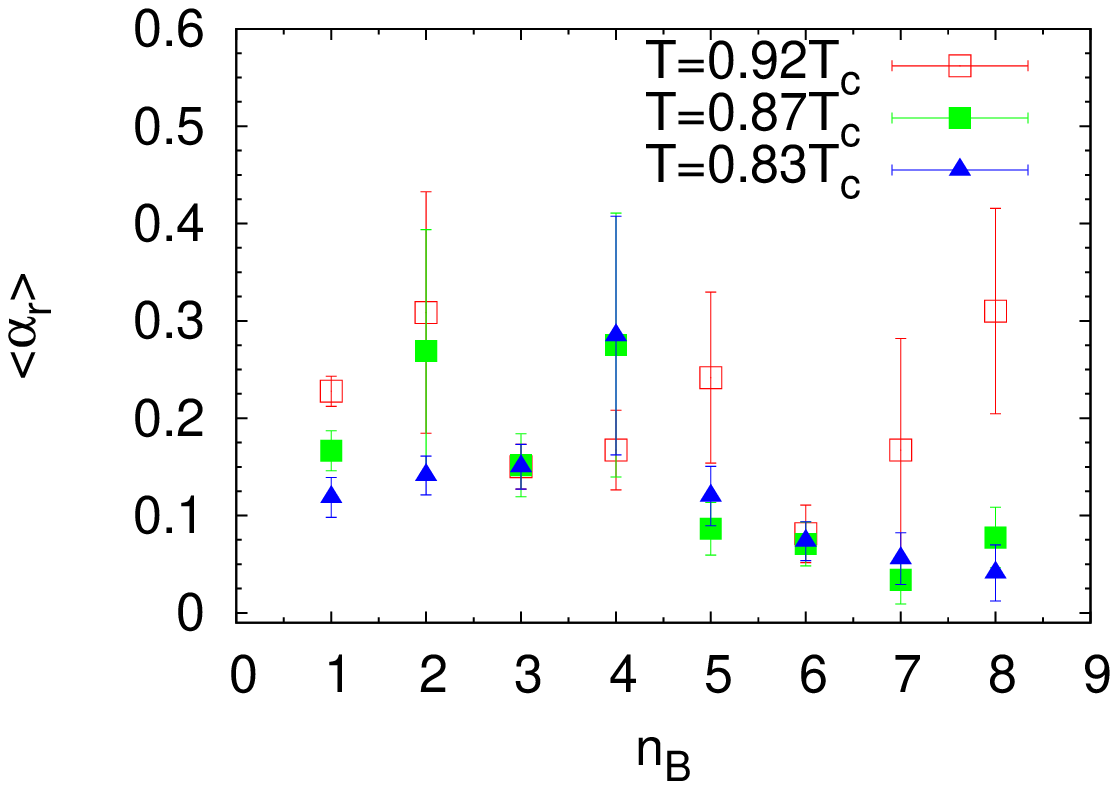}
\caption{Left: Average sign of $N_f=4$ in terms of $n_B$. Right: Average sign of $N_f=2$
in terms of $n_B$}\label{fig:sign_nf24}
\end{figure}

As a first check, we examine the seriousness of the potential sign problem in the canonical approach. The average sign in Eq.~(\ref{eq:sign_rew})
is plotted in Fig.~\ref{fig:sign_nf24}. For $N_f=4$ (left panel), 
we see that they range mostly from 0.6 to 0.2. Except for the point at $T=0.89~T_c$ and $n_B=9$ which
has a 1.5 sigma away from zero, the others
all have more than 3 sigmas above zero. Again, for $N_f=2$ (right panel), we see that except for the case at the higher $n_B$ and lower temperature, the average sign for most of the cases is more than three sigmas above zero. In view of the fact that the results we are going to show agree with those from the
imaginary chemical potential study and 
the extrapolated boundaries at finite density on $N_f=4$ meet at the expected $T_c$ at $\mu =0$, we believe that the sign
fluctuation is not a problem.

\subsection{Baryon chemical potential}

Before starting the discussion on the QCD phase diagram, we will first
present the ``tool'' we used to determine the phase transition in the
canonical ensemble: the baryon chemical potential. We measure the chemical potential
and plot it as a function of the net quark number $k$.
Due to the contribution of the surface tension to the free energy at finite volume,
the chemical potential in the mixed phase region will display an ``S-shape'' structure.
We will scan the baryon number at fixed temperature to look for this S-shape wiggle in
chemical potential which is a signal for a first order phase transition.

In the canonical ensemble, the baryon chemical potential is calculated by taking the difference of the
free energy after adding one baryon~\cite{Alexandru:2005ix}, i.e.
\begin{equation}
\left<\mu\right>_{n_B} = \frac{F(n_B+1)-F(n_B)}{(n_B+1)-n_B} = -\frac{1}{\beta}\ln
\frac{Z_C(3n_B+3)}{Z_C(3n_B)} = -\frac{1}{\beta}\ln \frac{\left<\gamma(U)\right>_o}{\left< \alpha(U)\right>_o}
\label{eq:baryon chemical potential}
\end{equation}
where
\begin{eqnarray}
\alpha(U) &=& \frac{\Re\detn{3n_B} M^{n_f}(U)}{\left|\Re \detn{3n_B}
M^{n_f}(U)\right|}, \quad {\mbox{and}} \\
\gamma(U)&=&  \frac{{\Re\,}\detn{3n_B+3} M^{n_f}(U)} {\left|{
\Re\,}\detn{3n_B} M^{n_f}(U)\right|}.
\end{eqnarray}
are measured in the ensemble with $n_B$ baryon number where $\left<\right>_o$ in
Eq.~\eqref{eq:baryon chemical potential} stands for the average over the ensemble generated with the measure
$\left|\Re \detn{3n_B}M^{n_f}(U)\right|$.

\subsection {QCD phase diagram}

\begin{figure}
\centering
\includegraphics[scale=0.4]{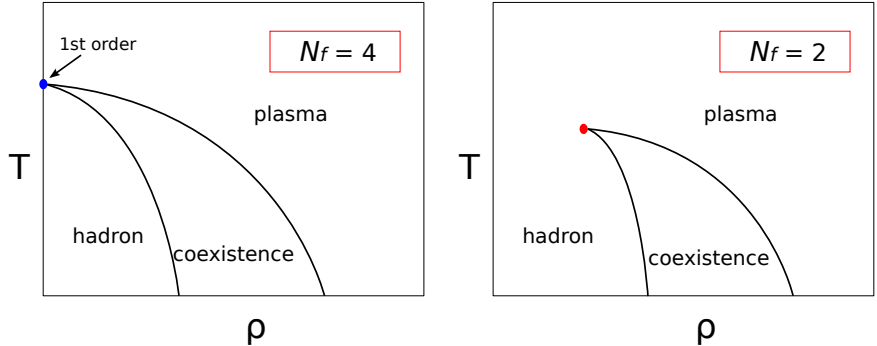}
\caption{Schematic phase diagram of four and two flavors in canonical ensemble.}\label{fig:nf2nf4_phase}
\end{figure}

Before we present our results, we would like to point out the difference between the phase diagram in
the grand canonical ensemble and the one in the canonical ensemble. We plot the expected canonical ensemble
phase diagram in Fig.~\ref{fig:nf2nf4_phase}. In the $T$ - $\rho$ diagram, the first order phase
transition line in the grand canonical $T - \mu$ diagram becomes a phase coexistence region which has two boundaries that separates
it from the pure phases. The two boundaries will eventually
meet at one point. For $N_f=4$, this point should be located at $\mu_B=0$ and the
transition temperature $T_c$.
For the $N_f = 2$ and $N_f = 3$ cases, if there are first order phase transitions, this
point will be the critical point at nonzero baryon chemical potential. Our method determines the position of the boundaries of the coexistence region at the temperatures
we use in our scanning; we use an extrapolation to locate the intersection point.

The reasons we study the two and four flavors systems are the following:
first of all, it is easier to simulate an even number of flavors with HMC~\cite{Duane:1987de}.
Secondly, in the four-flavor case, the first order phase transition extends all the way to zero density. 
We can then test our method by extrapolating the non-zero density results back to zero density to check if the transition point identified this way agrees with that determined from the the peak of the Polyakov loop susceptibility. Moreover, we can compare our results with those from simulations using imaginary chemical potential or those relying on a $\mu/T$
Taylor expansion; these simulations are expected to identify the transition point reliably for small 
baryon densities.
Thirdly, the two-flavor case is expected to have a phase diagram similar to that of
real QCD (see Fig.~\ref{fig:nf2nf4_phase}).

For the above reason, we will use the
four- and two-flavor
simulations as a benchmark to demonstrate the
methodology of determining the phase boundaries and locating the critical point before tackling the more realistic case.

We start by determining the location of the transition to the quark-gluon plasma phase at zero
chemical potential. We determine the order of the phase transition as well as the
location of $T_c$ using the Polyakov loop susceptibility:
\beq
\chi_P=VN_t\left<(P-\langle P\rangle)^2\right>
\eeq
In a finite volume, the susceptibilities are analytic functions, even in the regime where phase transitions
occur. In the infinite volume limit, on the other hand, phase transitions reveal themselves through the divergences of susceptibilities; for crossovers,
susceptibilities remain finite. The order
of the transition can be determined by finite size scaling of the susceptibilities. The susceptibility at
peak $\chi_{peak}$ behaves as $\chi_{peak}
\propto V^\alpha$, with $\alpha$ being the critical exponent. If $\alpha=0$, the transition is just a crossover;
if $0 <\alpha<1$, it is a second order phase
transition and if $\alpha=1$, it is a first order phase transition. 
We run simulations for two different
volumes $6^3\times4$ and $10^3\times4$ and found $\alpha=0.45(3)$ for $N_f=2$ and $\alpha=1.01(3)$ for $N_f=4$.
It is clear that, the $N_f=4$ case has a
first order phase transition at zero chemical potential which is consistent with other calculations at larger volumes~\cite{Fukugita:1990vu} as well as a theoretical prediction~\cite{Pisarski:1983ms}.

\begin{figure}[hbt!]
\centering
      \includegraphics[scale=0.30]{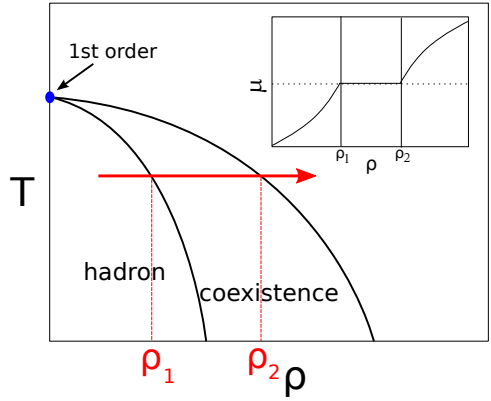}
      \caption{Schematic plot illustrating the scanning we use to locate the
boundaries of the mixed phase for QCD with  $N_f=4$. The infinite volume
expectation for chemical potential as a function of
density is shown in the insert.} \label{fig:t_rho_scan}
\end{figure}

We scan the phase diagram by fixing the temperature below $T_c$  while varying $n_B$.
Once we enter the coexistence region in a finite volume, the contribution from the surface
tension causes the appearance of a ``double-well'' effective free energy which
leads to an S-shaped behavior in the chemical potential vs. baryon number plot~\cite{Ejiri:2008xt}.
In the thermodynamic limit, the surface tension contribution goes away since it is a surface
term and the free energy scales with the volume;
the chemical potential will then stay constant in the mixed-phase region.
We sketch the expected $N_f =4$ phase diagram and mark the boundaries $\rho_1$ and
$\rho_2$ at a temperature below $T_c$ in  Fig.~\ref{fig:t_rho_scan}. The baryon chemical potential
in the thermodynamic limit is shown in the insert.

\begin{figure}[hbt!]
\centering
      \includegraphics[scale=0.49]{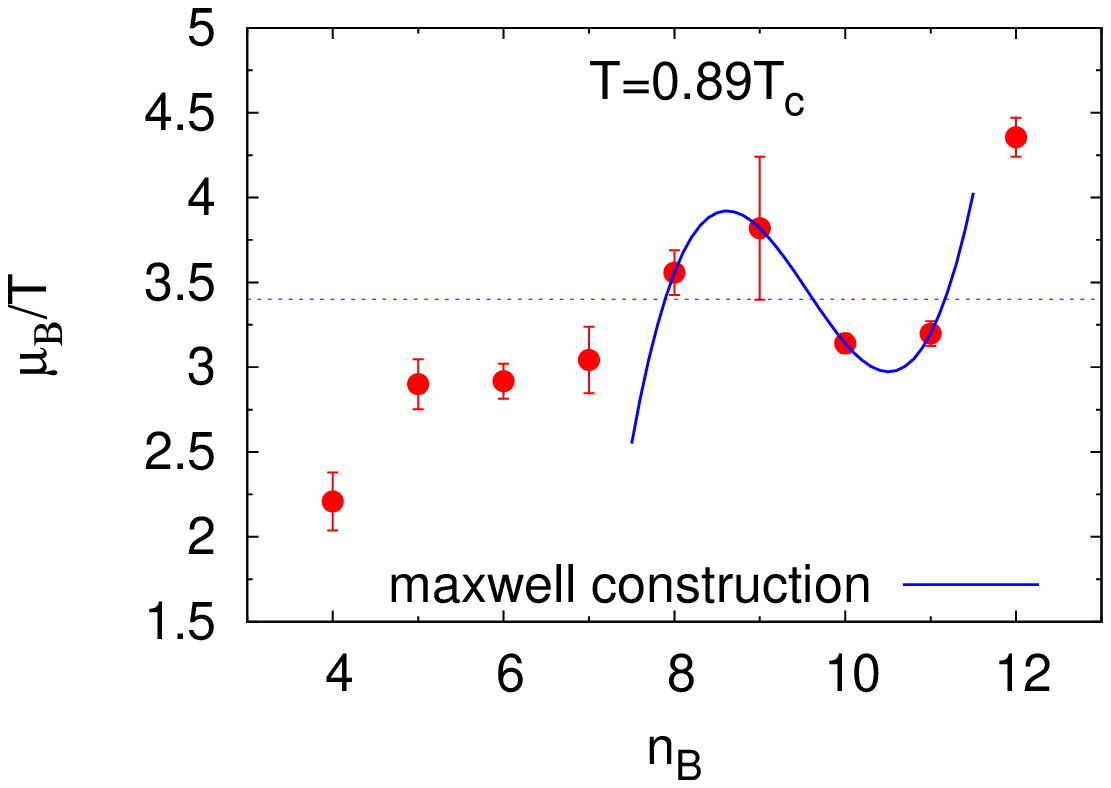}
      \includegraphics[scale=0.49]{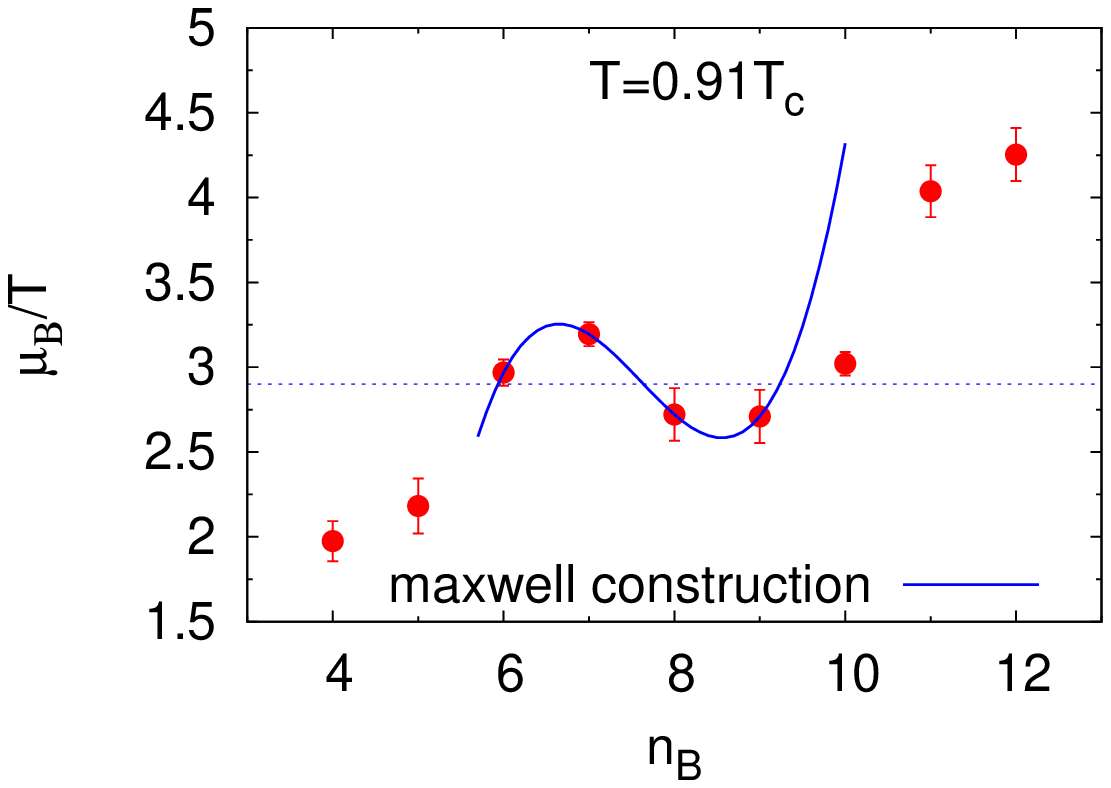}
      \includegraphics[scale=0.49]{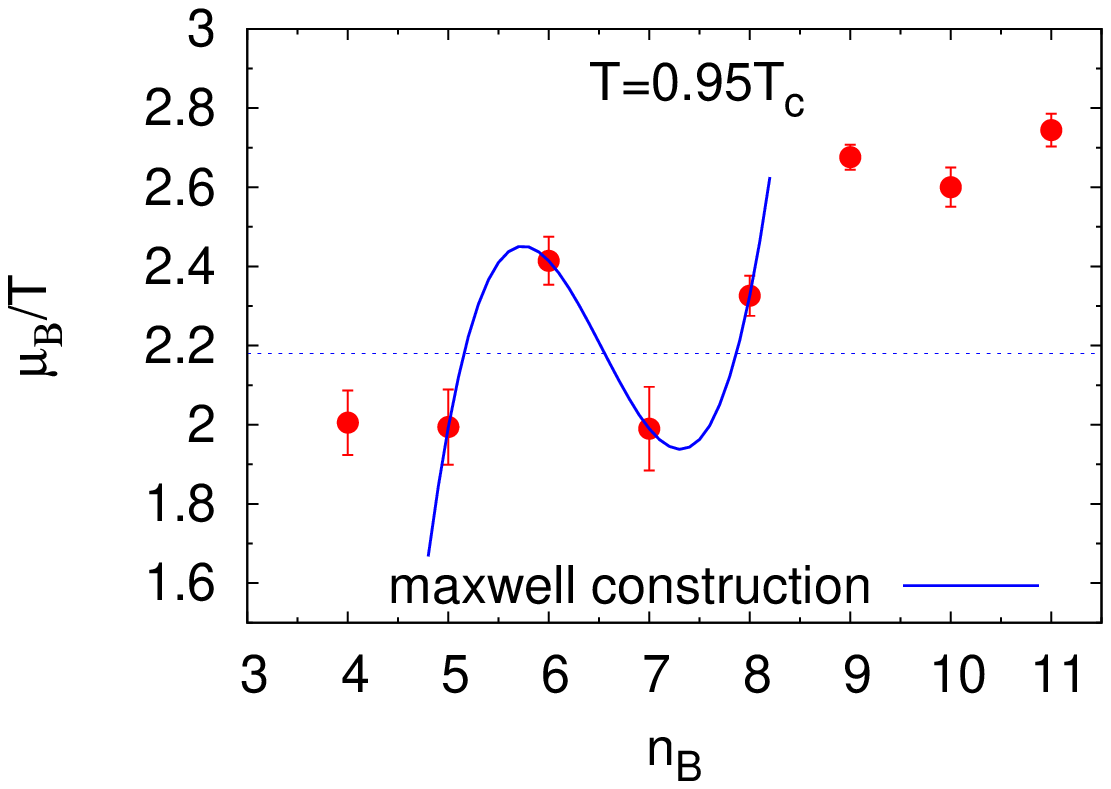}
\caption{$N_f=4$ phase scan along with the dashed line indicating the coexistence chemical potential $\widetilde{\mu_B}/T$ 
computed using the Maxwell construction.}\label{fig:scaning_nf4}
\end{figure}

\begin{figure}[hbt!]
\centering
      \includegraphics[scale=0.7]{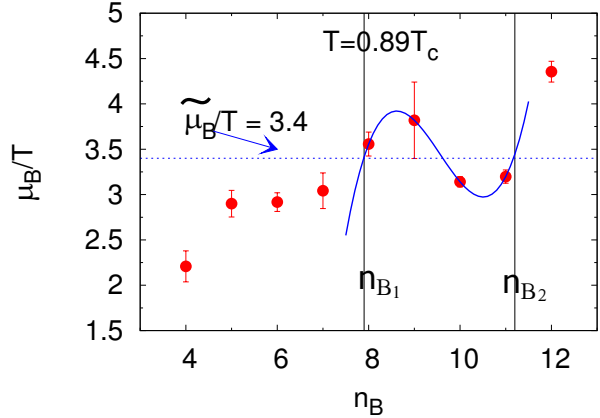}
\caption{Maxwell constructions for $T=0.89~T_c$ with the horizontal dashed line indicating the constant $\widetilde{\mu_B}/T$ and
the vertical lines indicating the mixed phase boundaries at $n_{B_1}$ and $n_{B_2}$.}\label{fig:maxwell}
\end{figure}

Our results for the baryon chemical potential for $N_f=4$ are presented in 
Fig.~\ref{fig:scaning_nf4} for three different temperature below $T_c$.
It is clear that the chemical potential exhibits
``S-shaped'' wiggles for $n_B$ between 5 and 11.
To identify the boundaries of the coexistence region and the baryon chemical potential of the coexistence phase, we rely on the ``Maxwell construction'': referring to Fig.~\ref{fig:scaning_nf4}
and Fig.~\ref{fig:maxwell}, the coexistence chemical potential $\widetilde{\mu_{B}}$ is the one that produces equal areas between the
curve of the chemical potential $\mu_B$ as a function of $n_B$ and the constant $\widetilde{\mu_B}$ line which
intersects with $\mu_B$ at $n_{B_1}$ and $n_{B_2}$. This procedure was first used in this context by 
de Forcrand and Kratochvila~\cite{deForcrand:2006ec}.

Since we do not have a functional description of how the chemical potential depends on the density
as we cross the coexistence region,
we select several points in the S-shape region and fit them with a third-order
polynomial. This approximation produces good results since the value of the coexistence chemical potential, 
$\widetilde{\mu_B}$, and the boundary points $n_{B_{1}}$ and $n_{B_{2}}$ are fairly insensitive to the
fitting function and the fit region; the simple third order polynomial fit is sufficient.

\begin{figure}[hbt!]
\centering
\includegraphics[scale=0.7]{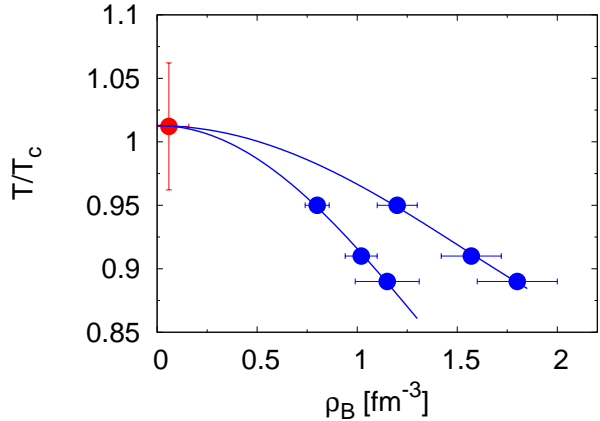}
\caption{Phase boundaries in the temperature vs.\ density plot for
$N_f=4$.}\label{fig:bound_nf4}
\end{figure}

We perform the Maxwell constructions for the three temperatures we studied for 0.89 $T_c$, 0.91 $T_c$ and 0.95 $T_c$ and
the results are presented in Fig.~\ref{fig:scaning_nf4}.
Having determined $n_{B_1}$ and $n_{B_2}$ for three different temperatures, we plot the boundaries of the coexistence region
and perform an extrapolation in $n_B$ to locate the intersection of the two boundaries.
Although there is no ``critical point''
for the $N_f=4$ case, it is expected that
the two coexistence phase boundaries should cross at zero chemical potential and $T=T_c$.
To determine the crossing point, we fit the boundary lines using a simple even polynomial in baryon density
\beq
\frac{T_c(\rho)}{T_c(0)} = 1 - a(N_f,m_q)(V\rho)^2+O\left((V\rho)^4\right)
\label{eq:expansion}
\eeq
to do the extrapolation. The phase boundaries and their extrapolations are plotted in Fig.~\ref{fig:bound_nf4}.
We find the intersection point at $T_c(\rho)/T_c(0) = 1.01(5)$ and
$\rho=0.05(10)\fm^{-3}$ which is consistent with our expectation of $\rho=0$ and $T=T_c$.

We would like to point out that the critical temperature was determined using a different set of simulations.
We ran simulations at zero baryon density and monitored the
Polyakov loop susceptibility as we increased the temperature. The critical temperature was determined using the peak of the Polyakov loop susceptibility. The fact that 
the intersection
point is consistent with the critical temperature determined from a
set of different simulations
constitutes a verification for our method of locating the
crossing point of the mixed phase boundaries.

\begin{figure}[h!]
\includegraphics[scale=0.65]{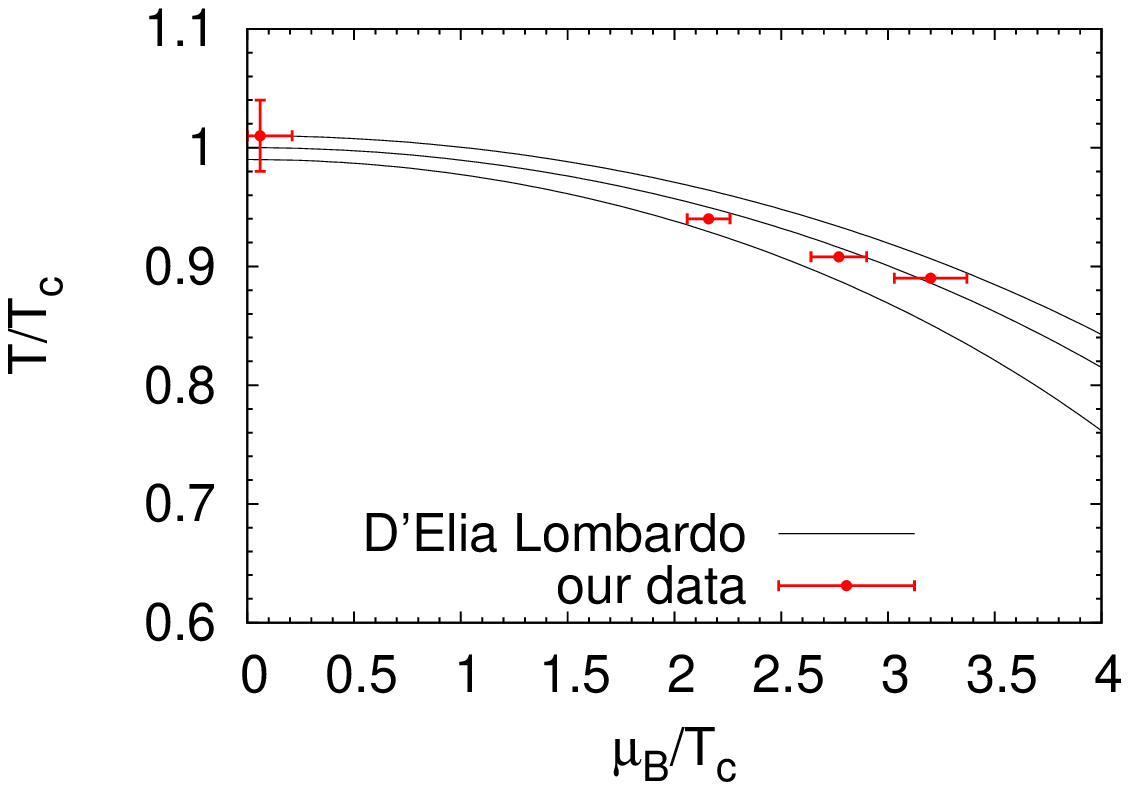}
\includegraphics[scale=0.27]{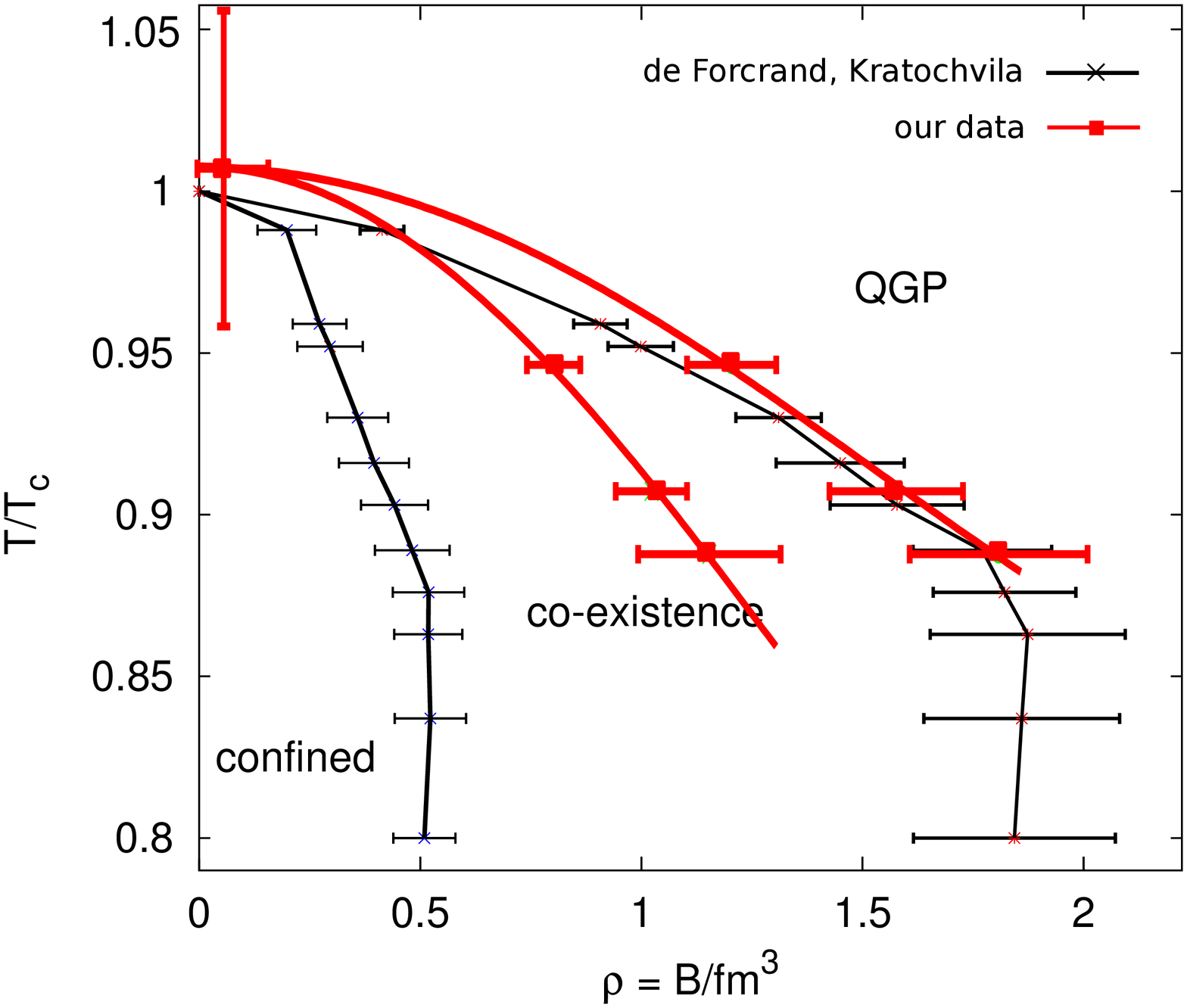}
\caption{Phase transition line for $N_f=4$ in the $T$, $\mu$ plane.}\label{fig:nf4_compare}
\end{figure}

We compare our results to those from a study based on an analytic continuation 
from the imaginary chemical potential with
staggered fermions~\cite{D'Elia:2002gd}.
The results are presented in Fig.~\ref{fig:nf4_compare} (left panel).
The solid curves represent the staggered fermion results with error band and our results are given
in solid circles~--~we see that the agreement\footnote{We also note the agreement with a recent study of $N_f=4$ via imginary chemical potential approach~\cite{Cea:2010md}.} is quite good in spite of the
fact that the quark mass used in the staggered fermion study, $m_\pi \approx 300-400\MeV$, is significantly
smaller than ours at $\sim 800$ MeV. 
This is not too surprising in view of the fact that the phase transition curve for $N_f=4$
is known to be fairly
independent of the quark mass~\cite{Philipsen:2008gf}.

We also compare our results to the ones from a canonical ensemble study that uses staggered
fermions~\cite{deForcrand:2006ec} (Fig.~\ref{fig:nf4_compare}, right panel).
We find that our coexistence region is narrower.
The upper boundaries between the mixed phase and the quark-gluon plasma phase seem to be 
consistent, but our lower boundary between the hadron phase and the mixed phase lies higher 
in baryon density than that found in the staggered fermion study.
The discrepancy could come from the difference in the quark masses. 
The quark mass in our study corresponds to $\sim 800\MeV$ pion mass, while 
for the staggered case it corresponds to $m_\pi\approx 300 \MeV$ .

\begin{figure}[hbt!]
\centering
\includegraphics[scale=0.49]{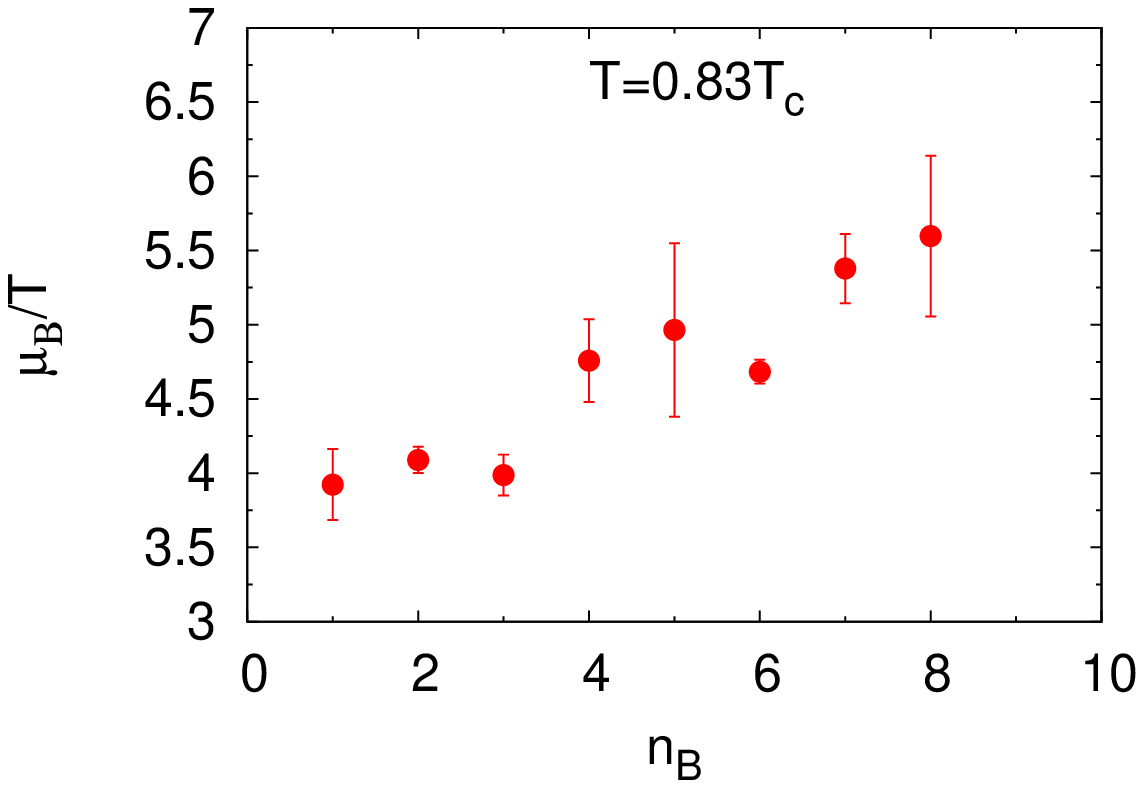}
\includegraphics[scale=0.49]{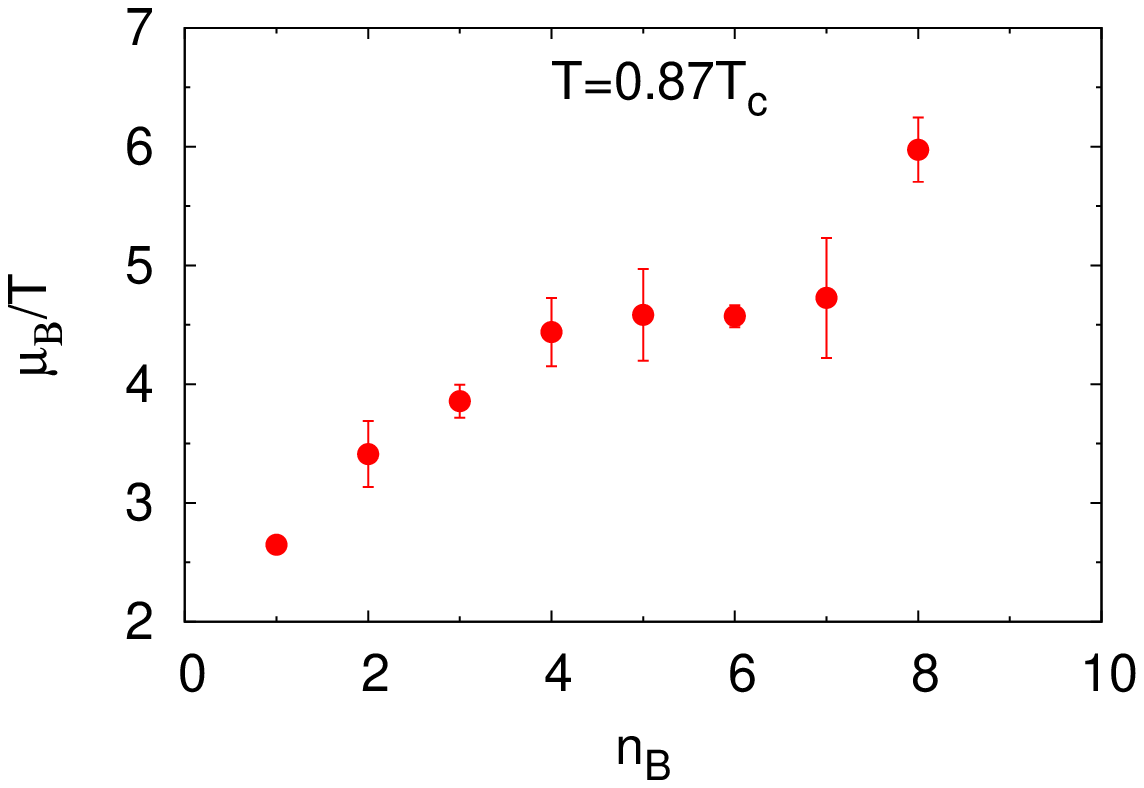}
\includegraphics[scale=0.49]{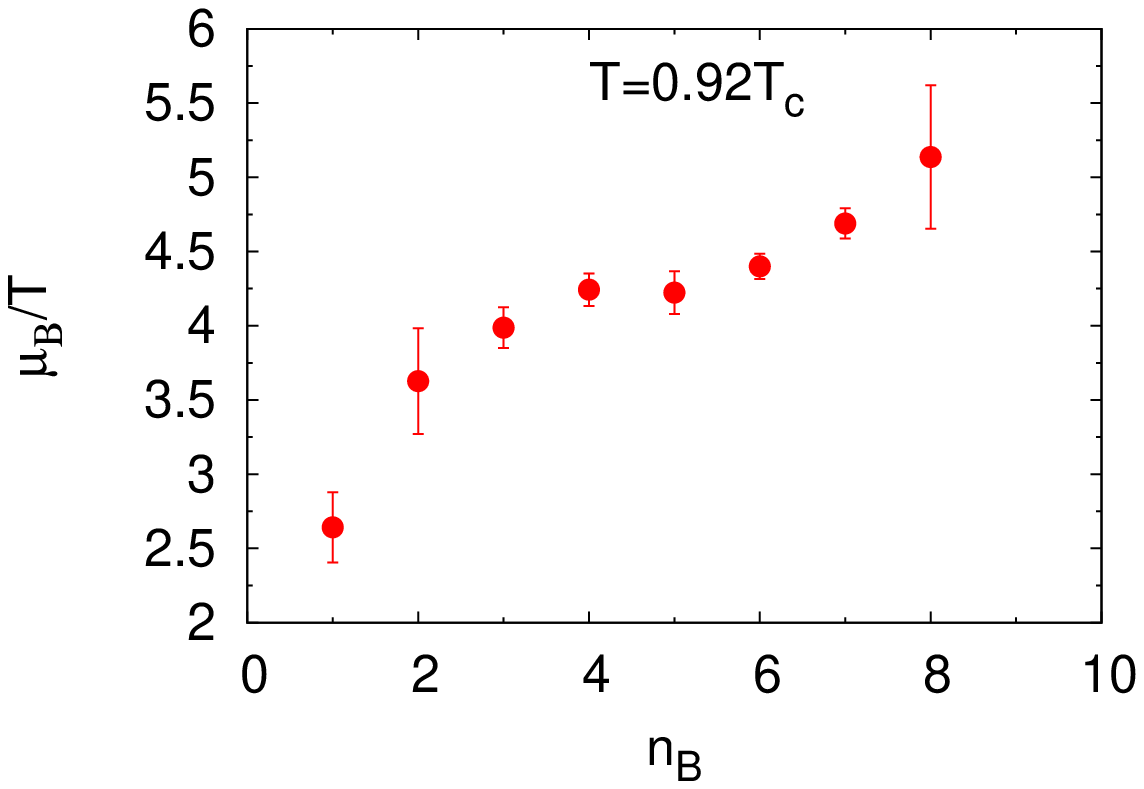}
\caption{Baryon chemical potential for $N_f=2$.}\label{fig:scaning_nf2}
\end{figure}

For the $N_f=2$ case, the phase diagram (see Fig.~\ref{fig:nf2nf4_phase})
is expected to be similar to the conjectured real QCD phase diagram.
For certain nonzero quark mass, it also
possesses
a crossover behavior at vanishing chemical potential, a critical point which ends the first order phase transition line.
For this system, we scan three temperatures below $T_c$ at 0.83 $T_c$, 0.87 $T_c$, and 0.92 $T_c$.
The results are presented in Fig.~\ref{fig:scaning_nf2}.
We do not observe a clear
signal for the S-shape structure within the scanned temperature and density range. This may be due to the fact that the
transition is milder than the sensitivity of our data, or it may be due to the
fact that the transition occurs at lower temperatures than the ones we studied -- there is at least one study that
claims that the critical point for $N_f=2$ occurs at temperatures below 0.8 $T_c$~\cite{Ejiri:2008xt}.

\section{Conclusion and outlook}

In this paper, we demonstrate that the canonical ensemble can be used to investigate QCD phase diagram at finite temperature
and nonzero baryon chemical potential.
The procedure we implemented to identify the first order phase transition
relies on scanning in baryon number at a fixed temperature and volume in order to search 
for the unique mixed phase signal in a finite volume as illustrated by an ``S-shaped''
structure. The well-established Maxwell construction is then utilized to determine
the boundaries between the mixed phase and the pure phases. Finally, the boundaries
so determined from several temperatures are extrapolated in baryon number and temperature 
to locate the intersecting point.

In order to scan the QCD phase diagram at large baryon number,
we have developed a winding number expansion method to calculate
the projected determinant.
The winding number expansion method expands the scanning region to 
baryon numbers as large as 12 in our study region without triggering
numerical instability which has rendered the discrete Fourier transform impractical.
We also note here that the properties of this projection were recently investigated by
a different group~\cite{Gattringer:2009wi,Danzer:2008xs}.

We presented our results for simulations of four and two flavors QCD. In the $N_f=4$ case, 
the first order phase transition at nonzero chemical potential appears as an
``S-shape'' structure in the plot of chemical potential vs. baryon number (baryon density). 
The ``S-shape'' structure is related to a finite volume effect which can be understood in terms
of a surface tension.
Although, to confirm the presence of a phase transition, the infinite volume limit needs to be
studied,
we interpret the appearance of the ``S-shape'' in the finite volume
as a strong indication
of a first order phase transition in the thermodynamic limit.
Maxwell construction is used to determine the phase boundaries for the $N_f=4$ case
at three temperatures below $T_c$.

The intersecting point from the extrapolation of the two boundaries turns out to fall on the
expected first order phase transition point at $T=T_c$ and zero baryon density.
Furthermore, our results for the critical chemical potential agree with those from the imaginary
chemical potential study with staggered fermions.
These facts show convincingly that our method can be used to determine the location of
the critical point.

In the $N_f=2$ case, we do not see any signal
of first order phase transition for temperatures as low as 
0.83 $T_c$.

Our present study is carried out on
a small volume ($6^3 \times 4$) with lattice spacings $a = 0.27 - 0.32$ fm and
quark masses around that of the strange quark mass. 
To get closer to the physical point,
one needs to have lower quark masses, larger volumes and several lattice spacing to
extrapolate to the continuum limit. This is very challenging since
larger volumes will require larger baryon numbers for the same density and this
is likely to increase the severity of the sign problem and lower the Monte Carlo acceptance rate.
On the other hand, larger volumes will allow us to use
smaller quark masses 
which, in turn, may lower the baryon density for the mixed phase due to larger baryon 
sizes. Noise estimator for the fermion 
determinant~\cite{Alexandru:2007bb} or better algorithms may be needed to meet this
challenge. 
Our next goal is
to extend our present study to the $N_f=3$ case. Even
with small volumes and large quark masses, it is interesting
to check whether there exists a first order phase transition for this system and, if it exists, to locate
its critical point.

\begin{acknowledgments}
We would like to thank P. de
Forcrand, S. Ejiri, C. Gattringer, F. Karsch and M.-P. Lombardo for their useful discussions and
suggestions.
The work is partially supported by U.S. Department of
Energy (USDOE) Grants No. DE-FG05-84ER-40154. A. Li is also supported by DOE grant DE-FG02-05ER-41368. A. Alexandru is supported in part by U.S. Department of Energy under grant DE-FG02-95ER-40907 and in part by the George Washington University IMPACT initiative. We appreciate the hospitality of the Kevli Institute of Theoretical Physics in Beijing
where part of this work was carried out during a lattice workshop in July 2009. The calculations
were performed at Texas Advanced Computing Center (TACC) at the Univ. of Texas at Austin
and Center for Computational Sciences at the Univ. of Kentucky.
\end{acknowledgments}

\bibliography{myref}

\end{document}